\documentclass[singlecolumn]{emulateapj}
\usepackage{float,epsfig}\usepackage{natbib}
\newcommand{\cxou}{CXOU~J102843.0$+$682816}
\newcommand{\msun}{$M_{\odot}$}
\newcommand{\msfr}{$M_{\odot}$~yr$^{-1}$}
\newcommand{\ergl}{erg~s$^{-1}$}
\newcommand{\hih}{H{\sc i} hole}
\newcommand{\ic}{IC~2574}

\newcommand{\um}{$\mu$m}
\newcommand{\sgs}{supergiant shell}

\newcommand{\ha}{H$\alpha$}
\newcommand{\hi}{H{\sc i}}
\newcommand{\hii}{H{\sc ii}}
\newcommand{\cxo}{{\sl Chandra}}
\newcommand{\hst}{{\sl Hubble}}
\newcommand{\ros}{{\sl ROSAT}}
\newcommand{\galex}{{\sl GALEX}}
\newcommand{\sst}{{\sl Spitzer}}

\slugcomment{Submitted to Astrophysical Journal Letters}

\begin{document}

\title{X-Ray Emission from the Supergiant Shell in IC~2574}

\author{
Mihoko~Yukita\altaffilmark{1} and
Douglas~A.~Swartz\altaffilmark{2}
}

\altaffiltext{1}{Department of Physics \& Astronomy,
    University of Alabama, Tuscaloosa, AL, USA}
\altaffiltext{2}{Universities Space Research Association,
    NASA Marshall Space Flight Center, VP62, Huntsville, AL, USA}

\begin{abstract}

The M81 group member dwarf galaxy \ic\ hosts a supergiant shell of
 current and recent star-formation activity surrounding 
 a 1000$\times$500~pc hole in the ambient \hi\ gas distribution.
\cxo\ {\it X-ray Observatory} imaging observations reveal a luminous,
 $L_{\rm X} \sim 6.5\times10^{38}$~\ergl\ in the 0.3$-$8.0~keV band,
 point-like source within the hole but offset from its center
 and fainter diffuse emission extending throughout and beyond the hole.
The star formation history at the location of the point source 
 indicates a burst of star formation beginning $\sim$25~Myr ago and 
 currently weakening and
 there is a young nearby star cluster,  at least 5~Myr old, bracketing the
 likely age of the X-ray source at between 5 and $\sim$25~Myr. 
 The source is thus likely a bright high-mass X-ray 
 binary~---~either a neutron star or black hole accreting from an
 early B star undergoing thermal-timescale mass
 transfer through Roche lobe overflow.
The properties of the residual diffuse X-ray emission are consistent with
 those expected from hot gas associated with the
 recent star-formation activity in the region.

\end{abstract}


\section{Introduction}

IC~2574 is a low metallicity 
\citep{miller96,masegosa91,croxall09}, gas-rich, 
dwarf irregular galaxy located 4.0~Mpc distant
 \citep[][1\arcsec$=$19~pc]{karachentsev02} in the M81 group of galaxies.

The galaxy hosts numerous \hi\ holes and 
shells \citep{walter98,walter99}, one of which
 is located in the most prominent region of 
 current and recent star-formation and is near the outskirts of the galaxy.
This 1000$\times$500~pc supergiant shell
 has been studied intensively at 
 UV \citep{stewart00},
 optical \citep{pasquali08,weisz09},
 mid-infrared \citep{cannon05},
 and radio wavelengths \citep{walter98,walter99}.
Recent analysis of resolved stars imaged with \hst\ \citep{weisz09} shows that 
 the most significant episodes of star formation began nearly
 100~Myr ago and isolated bursts along the shell periphery
 are as young as 10~Myr. 
The ages of the younger star-formation events are consistent with
 those derived from broadband photometry
 \citep[e.g.,][]{stewart00,cannon05,pasquali08}
 and only slightly younger than the estimated dynamical age of the \hi\ shell,
 14~Myr \citep{walter98,walter99}.

\citet{walter98} also report analysis of soft X-ray emission from
 the \sgs\ observed with the \ros\ PSPC. 
They found that the emission was extended beyond the $\sim$28\arcsec\
 FWHM instrumental point-spread function (PSF) 
 with a surface brightness peak near the center of the \hi\ hole. 
Based on spectral hardness ratios, \citet{walter98} compute a
 plasma temperature of $\sim$0.54~keV, 
 a luminosity of 1.6$\times$10$^{38}$~\ergl\ in the 0.3$-$2.4~keV range,
 and an electron density of 0.03~cm$^{-3}$
 (assuming a metallicity $Z/Z_{\odot}=0.15$ and a spherical 
 emitting volume of 700~pc diameter).

Here, we analyze higher resolution \cxo\ spectrophotometric imaging 
 observations of the \sgs.
We show that the bulk of the emission arises from
 a bright point source located within the \hi\ hole (\S~\ref{s:ptsrc})
 with a mildly-absorbed hard power law spectrum (photon index 1.6)
 and intrinsic luminosity 6.5$\times$10$^{38}$~\ergl\ in the 0.3$-$8.0~keV range.
There remains residual X-ray emission (\S~\ref{s:diffx}) which can be 
 characterized as a diffuse thermal plasma  
 (temperature 0.5~keV, 0.3$-$8.0~keV luminosity 2.7$\times$10$^{37}$~\ergl).
The emission extends throughout and beyond the \hi\ hole and 
 is likely the result of the recent star formation activity.

We then determine the mass, age, and extinction towards several 
 young star clusters inside and on the \sgs\ (\S~\ref{s:mimfes})
and use this to argue that the bright point source is a
 high-mass X-ray binary (HMXB) with an early B star companion (\S~\ref{s:discuss}).
The residual X-ray emission accounts for no more than $\sim$1\% of the
 mechanical energy produced by massive stars and supernovae in the
 massive star cluster in the \hih.

\section{X-ray Image Analysis}

\begin{figure*}
\begin{center}
\includegraphics[angle=0,width=0.9\columnwidth]{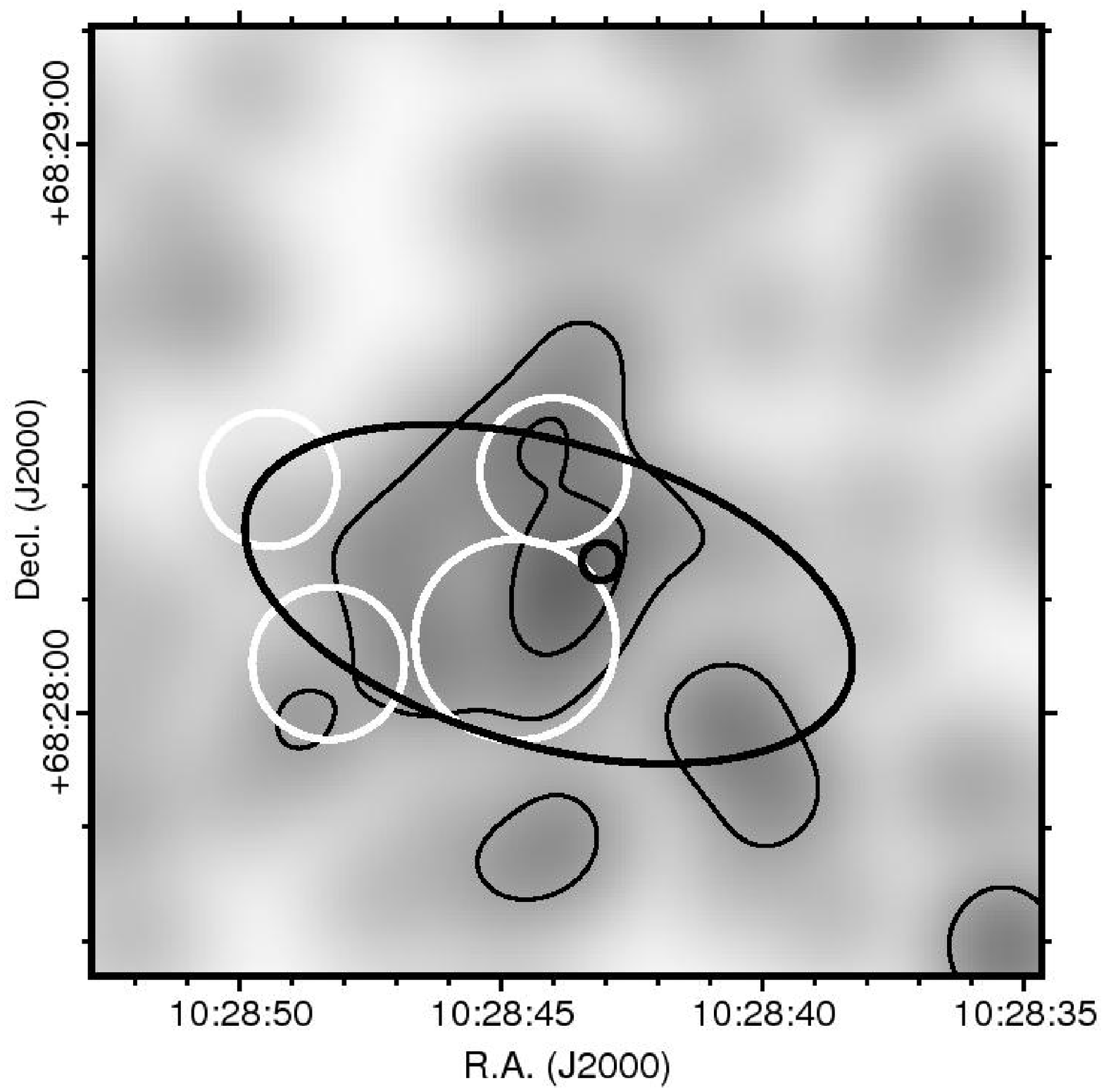}
\includegraphics[angle=-90,width=0.9\columnwidth,origin=c]{f2.eps}
\vspace{10pt}
\figcaption{{Left panel:} the  
diffuse X-ray emission in a 100\arcsec$\times$100\arcsec\ region
 of the supergaint shell of IC~2574.
The ellipse denotes the \hih\ as defined by \citet{walter98}
 and the small circle the position of \cxou.
The other symbols denote regions identified in Figure~\ref{f:ic2574smallsfregions}.
The contour levels are at 1 and 2$\times$10$^{-9}$~photon~cm$^{-2}$~s$^{-1}$~pixel$^{-2}$
(1~pixel$=$0\farcs492).
{Right panel:} deprojected radial profile of X-ray events 
 (unsmoothed; in the 0.3$-$2.0~keV energy range). 
The deprojection assumed that the elliptical \hih\ is circular.
 \label{f:ic2574sgscontours}}
\end{center}
\end{figure*}

\ic\ was observed with the \cxo\ {\it X-ray Observatory} 
 Advanced CCD Imaging Spectrometer (ACIS)
 on 2000 January 7 (ObsID 792) and again on 2008 June 30 (ObsID 9541).
Both observations were of approximately 10 ks duration 
 and are sensitive to on-axis point-like sources as faint as 
 $\sim$1.5$\times$10$^{37}$~\ergl.
Here, the earlier observation is analyzed because the 
 \hih\ was imaged close to the aimpoint and
 the back-illuminated S3 CCD used for this observation
 has higher effective area
 below 1~keV, where most of the X-ray
 emission is expected from a hot gas bubble,
 than the front-illuminated ACIS-I array used for the latter observation. 
We also note that the observation in 2000 was taken early enough in the mission
 that the soft response of the ACIS CCDs was not yet compromised by the buildup of
 material on their Optical Blocking Filters.
The observation was made in the full-frame timed exposure mode using the
 standard 3.2~s frame time and the VFAINT telemetry format.

The data were reprocessed
 beginning with the Level 1 event list.
A time-dependent gain correction was applied
 and pixel randomization was removed using the \cxo\ X-ray Center's
 CIAO ver. 4.3 tool {\tt acis\_process\_events}. 
The standard {\sl ASCA} event grades were selected
 and hot pixels, columns, and cosmic-ray afterglows were
 removed to create a Level 2 event list for analysis.
An $\sim$800~s period of high background near the end
 of the observation
 was omitted leaving a usable exposure time of $\sim$9~ks.

\section{CXOU~J102843.0$+$682816} \label{s:ptsrc}

There is a bright X-ray source located inside the supergiant shell.
\citet{walter98} judged it to be extended based on 
 a {\it ROSAT} PSPC observation.
However, it is clear in the \cxo\ image that most of the X-ray
 emission is contributed from a single point-like source.  
An elliptical Gaussian model (plus a constant term) 
 was fit to the spatial distribution
 of X-ray events in the 0.3$-$8.0~keV range to determine the 
 source centroid, Gaussian width, and approximate source asymmetry.
The best-fitting source position is 
 10$^{\rm h}$28$^{\rm m}$43.08$^{\rm s}$, $+$68\arcdeg28\arcmin16.37\arcsec.
The source is symmetrical within measurement errors with an rms
 Gaussian width of 0.315\arcsec$\pm$0.004\arcsec.

A radial profile of the data was compared to a simulation of the PSF built using the
 \cxo\ Ray Tracer (ChaRT) and MARX suites of programs available from
 the \cxo\ X-ray Center.
Fitting a Gaussian to the profile of the data and of the simulated PSF
 (plus a Lorentzian for the wings and a constant for the background)
 results in a best-fit width of the simulated PSF slightly larger than the observed 
 source width.
The source is clearly point-like.

Events were extracted from a 3\arcsec\ radius region surrounding the
 source for spectral analysis. This encircles roughly 99\% of the 
 0.3$-$8.0 keV flux from the source. 
 A background spectrum was extracted from the surrounding region.
An absorbed power law model, including a multiplicative pileup correction, was fit to the 
 background-subtracted spectrum using the XSPEC ver. 11.3.2ag analysis tool.

The best-fitting model column density is
 (5.4$\pm$5.8)$\times$10$^{20}$~cm$^{-2}$
 which is roughly the Galactic line-of-sight
 value of 2.2$\times$10$^{20}$~cm$^{-2}$,
 the best-fitting power law index is 1.6$\pm$0.6
 ($\chi^2=20.1$ for 31 dof),
 and the intrinsic X-ray luminosity is (6.5$\pm$1.5)$\times$10$^{38}$~\ergl\
after correcting for $\sim$10\% pileup.
A thermal plasma model also provides an acceptable fit to the point
 source spectrum but the resulting temperature, $kT_e \sim 5$ to 18~keV,
 and X-ray luminosity are much too high to be representative of
 a hot gas phase associated with star formation.

\begin{deluxetable*}{lcccrrr}
\tablecolumns{7}
\tablewidth{0pt}
\tablecaption{Properties of Star-forming Regions \label{t:ic2574sfrsed}}
\tablehead{
 \colhead{Region}  &  
 \colhead{Age}  & 
 \colhead{Mass} & 
 \colhead{$A_{V}$} & 
 \colhead{$L_{\rm W}$} &
 \colhead{Net Counts}&
 \colhead{$L_{\rm X}^{\rm int}$ }
\\
 \colhead{}  &  
 \colhead{(Myr)} & 
 \colhead{($10^{5}$~\msun)} &
 \colhead{(mag)} &
 \colhead{(10$^{38}$~erg~s$^{-1}$)} & 
  \colhead{} &
 \colhead{(10$^{35}$~\ergl)}}
\startdata
C1 &  7  &  0.8 & 0.0   & 27  & 6.8$\pm$2.8 & 38$\pm$16\\
C2 &  17 &  1.5 & 0.0   & 29  &9.3$\pm$3.3  & 52$\pm$19\\
C3 &  17 &  0.9 & 0.1   & 16   &1.0$\pm$1.4 &  6$\pm$8  \\
C4 &  7  &  0.2 & 0.1   & 4.7   &$-$0.4$\pm$1.0& $-$2$\pm$6\\
\enddata
\end{deluxetable*}

\section{Diffuse X-ray Emission} \label{s:diffx}

There remains an excess of X-ray events within and around the \hi\ hole (66\arcsec$\times$32\arcsec)
 after removing \cxou\ from the image.
This soft X-ray diffuse emission has low surface brightness,
 which leads to a grainy, low S/N, image.
The emission can be seen more easily if the image is artificially smoothed. 
The point source 
 was removed using a 3\arcsec\ radius circle, which is large enough to avoid 
 leaving residual emission from the wings of the point source.
Then, the excluded region was filled using the local background
 level  Poisson method of the CIAO tool {\tt dmfilth}.  
The filled image was then divided by a 0.5~keV monochromatic exposure map, 
 then smoothed with the CIAO tool {\tt aconvolve} using a 
 Gaussian smoothing function with a width of 10 pixels ($\sim$5\arcsec).

The left panel of Figure~\ref{f:ic2574sgscontours} shows the smoothed soft
 X-ray diffuse map of a 100\arcsec$\times$100\arcsec\ region
 around the supergiant shell.
The \hih\ is denoted by an ellipse and the location of \cxou\ by a small circle.
There is a net of 28.1$\pm$5.3 X-ray events in the energy range of 0.3$-$8.0~keV inside the
 \hi\ hole after subtracting background.
(The background was defined as emission in the S3 chip outside the optical extent
 of \ic, defined by an ellipse approximating the 25~mag~s$^{-1}$ blue 
 light isophote, after excluding any detected point sources.) 
We note that 90\% of the emission is found in the soft band (0.3$-$2.0~keV).

The right panel of Figure~\ref{f:ic2574sgscontours} displays the
 radial profile of events in the 0.3$-$2.0~keV energy range
 of the diffuse emission after deprojecting the image.
The deprojection was made assuming that the 66\arcsec$\times$32\arcsec\
 elliptical \hi\ hole  is actually circular.
The radial profile is centered on the \hi\ hole center.
It clearly shows the excess emission within the \hi\ hole and 
 perhaps extending slightly beyond the hole.
Note that there is no emission enhancement within the 
 current star-forming regions in the shell surrounding the hole.
The broad energy distribution of the X-ray events within the \hih\ was compared to
 a grid of models of an absorbed thermal plasma.
Specifically, the ratios of observed events in the 0.3$-$1.0 to 0.3$-$8.0 and in the
 0.3$-$1.0 to 0.3$-$2.0 bands most closely match a model of temperature
 0.5$\pm$0.1~keV and intervening absorption column
 of (4$\pm$2)$\times$10$^{20}$~cm$^{-2}$ where the errors denote the model grid spacing.
This model results in a luminosity of (2.7$\pm$0.5)$\times$10$^{37}$~\ergl\
 in the 0.3$-$8.0~keV range where the error is based on the net source counts.
 
\section{Mass, Age, and Extinction in Supergiant Shell Star-Forming Regions}
 \label{s:mimfes}

As mentioned in the introduction,
 several previous studies have provided estimates of 
 the mass and/or age of star-forming regions associated with the \sgs.
We have performed our own multi-wavelength analysis which produces
 self-consistent age, mass, and extinction estimates for individual
 (assumed co-eval) star clusters.
The method is presented in detail in \citet{yukitaD}.

Briefly,
the analysis uses the Starburst99 stellar
 synthesis program \citep{leitherer99} to 
 compute star cluster model spectra as a parameterized
 function of cluster age, star formation history, and metallicity
 from a combination of stellar evolution tracks and
 a grid of stellar model atmospheres weighted by an initial mass function (IMF).
We assume a (single) instantaneous burst history, solar metallicity,
 and a Salpeter IMF \citep{salpeter55}.
Transmission of the composite stellar spectra
 through the surrounding ISM is modeled 
 using a standard starburst dust extinction model \citep{calzetti00}
 that modifies the blue
 portion of the spectrum and we assume that the dust re-emits 
 this radiation as a blackbody 
 \citep[fixed at $T=30$~K, see][]{cannon05} in the IR band. 
Finally, the (dust-modified) 
 emergent model spectra are convolved with instrumental
response functions appropriate to the various instruments and 
 the best-fitting solution to the observed broadband photometry
 is determined. 
(The observations are corrected for a line-of-sight 
 Galactic extinction of $A_V$ = 0.120 \citep{schlegel98} 
 using the \citet{cardelli89} dust model.)
This method self-consistently overcomes
 the age-extinction degeneracy when fitting models
 to UV-to-IR broadband photometry.  

\begin{figure*}
\begin{center}
\includegraphics[angle=-00,width=0.9\columnwidth]{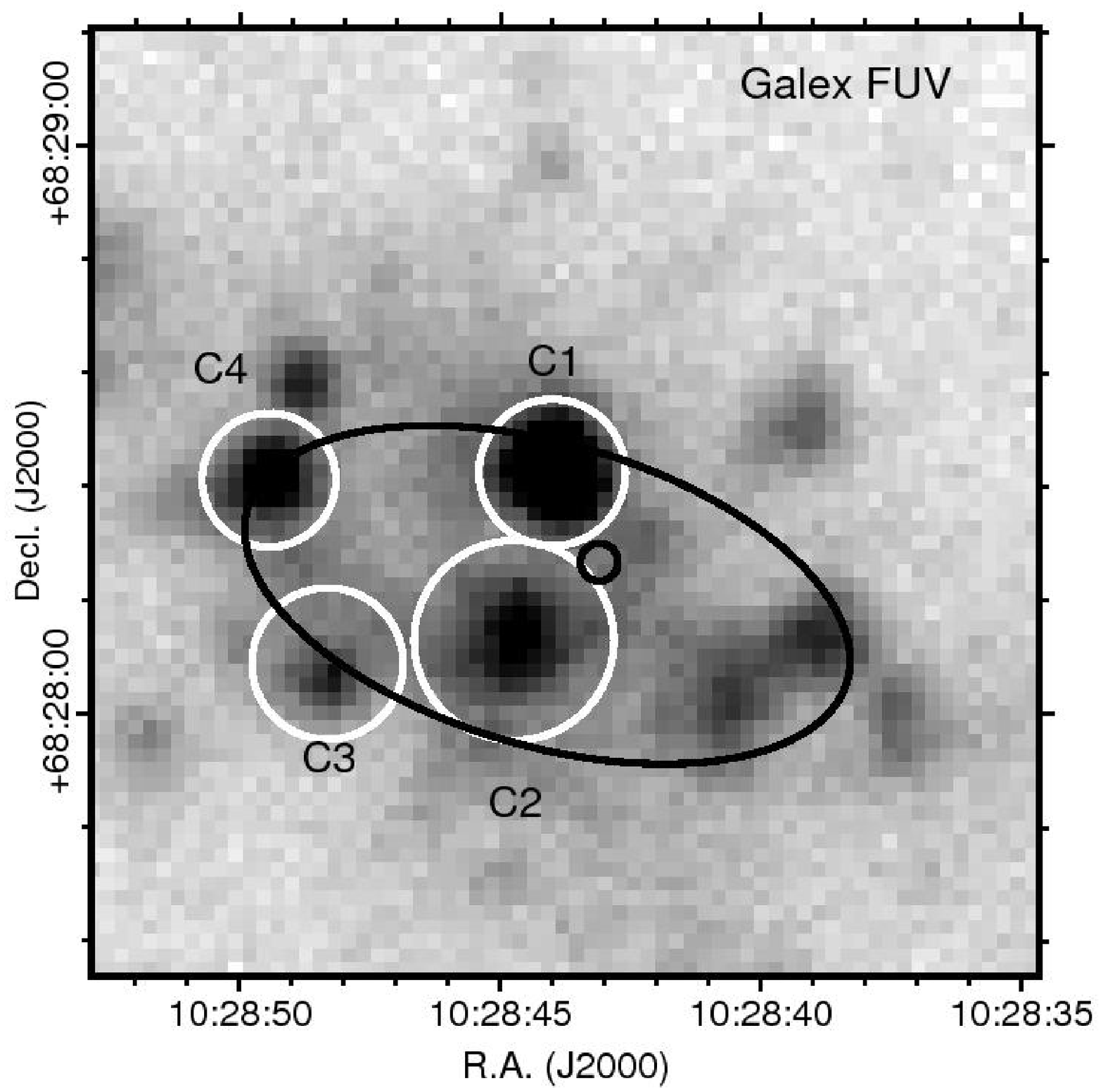}
\hspace{0.1in}
\includegraphics[angle=00,width=0.9\columnwidth]{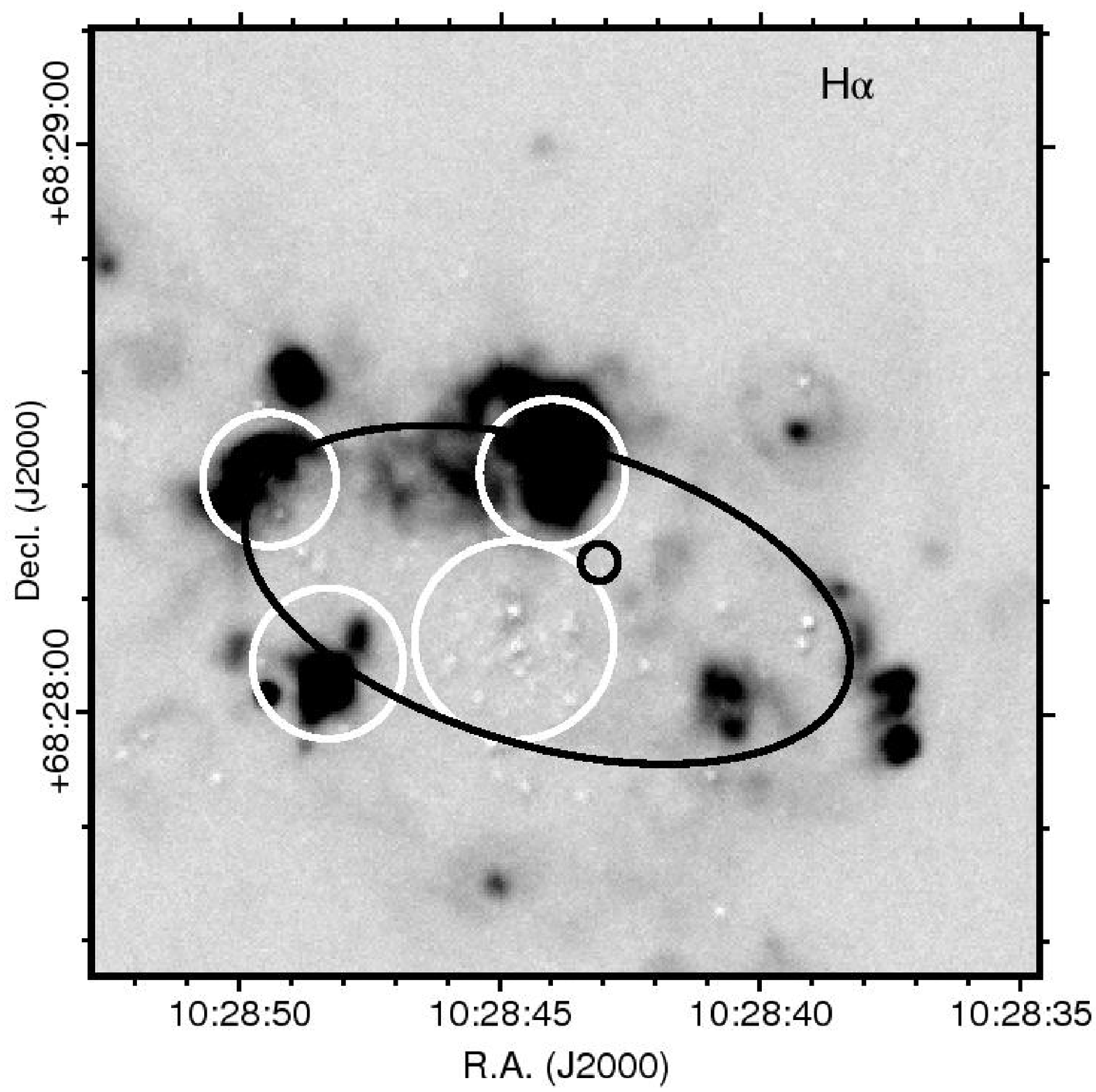}
\includegraphics[angle=-00,width=0.9\columnwidth]{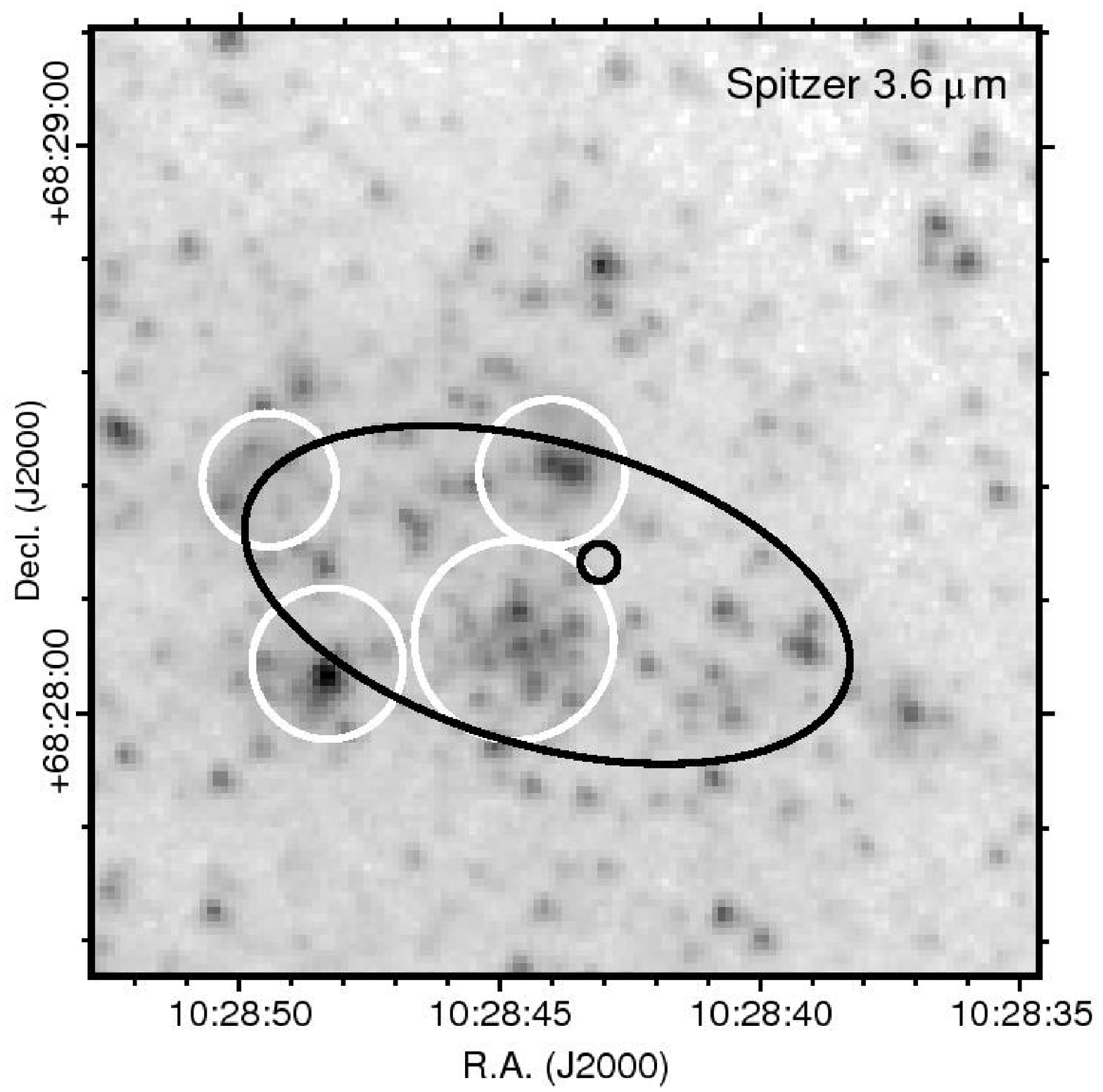}
\hspace{0.1in}
\includegraphics[angle=-00,width=0.9\columnwidth]{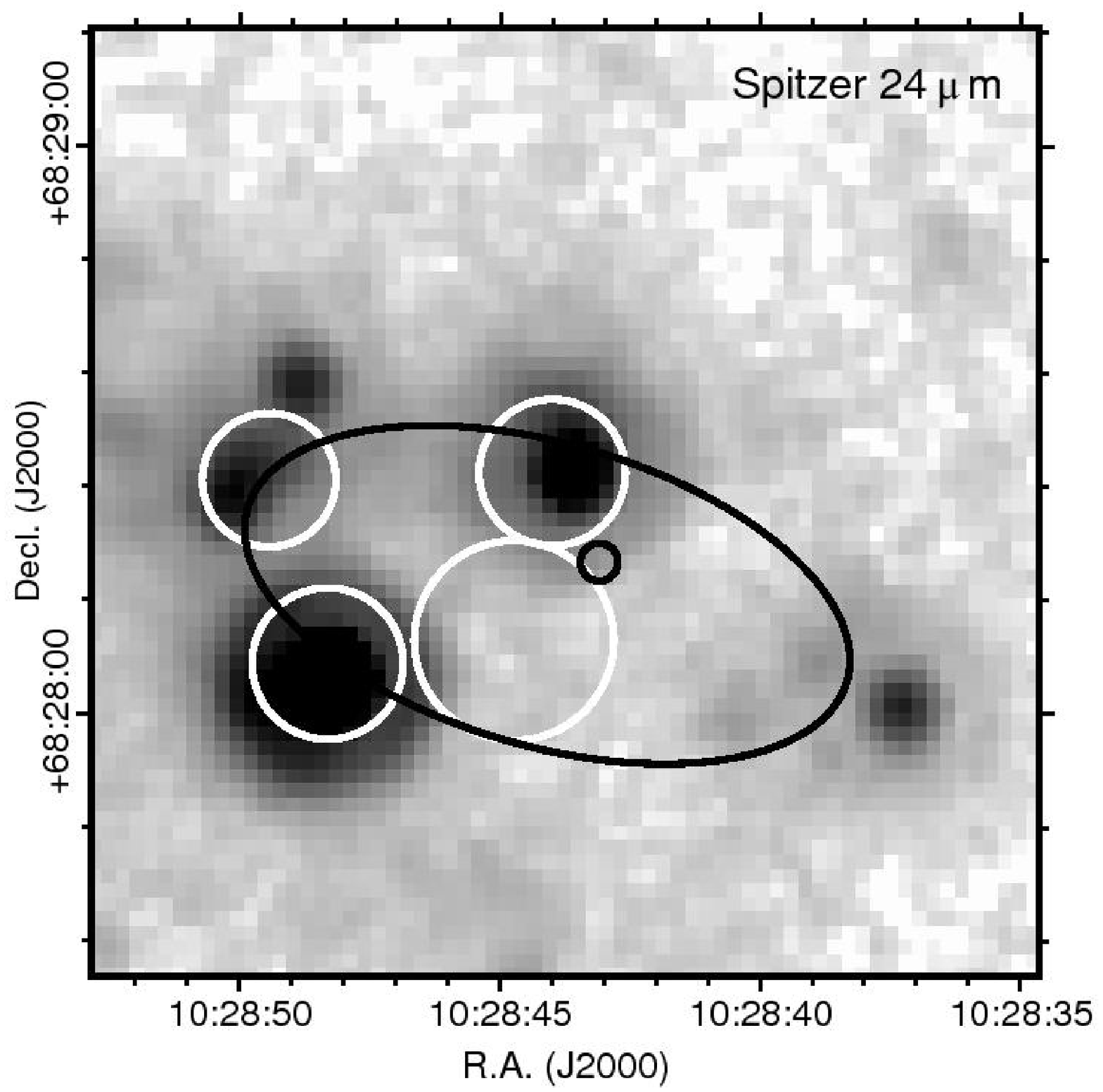}
\vspace{10pt}
\figcaption{Clockwise from top left: FUV, \ha, 24\,\um,
 and 3.6\,\um\ images of the same \sgs\ region as depicted in
 Figure~\ref{f:ic2574sgscontours}.
Four star clusters identified in the 24~\um\ and FUV
 are denoted by white circles depicting 3$\sigma$ aperture
 sizes defined by circular Gaussian fits to the FUV data.
\label{f:ic2574smallsfregions}}
\end{center}
\end{figure*}

Here, we use calibrated \galex\, FUV and NUV observations available through the 
 {\tt GalexView} data interface\footnote{http://galex.stsci.edu/GalexView/} 
 and \sst\ 3.6, 4.5, and 24 \um\ observations processed and made available
 as part of the \sst\ Infrared Nearby Galaxies Survey \citep{kennicutt03}.
Images of the \sgs\ at several passbands are displayed in 
 Figure~\ref{f:ic2574smallsfregions}. 
A continuum-subtracted \ha\ image observed with the Kitt Peak 2.1~m 
 telescope using the T2KB CCD Imager in 2008 March
 (1.3\arcsec\ seeing) is also shown.

Our cluster selection is to be both FUV and 24~\um\ bright sources.
Three clusters around the shell were identified from both  FUV and 24~\um\
 images. 
The UV bright cluster located in the \hi\ hole has been well studied \citep{stewart00,pasquali08} 
 therefore it is also investigated further, although it 
lacks 24~\um\ emission.
Overall, the total of four isolated star clusters were analyzed.
These are marked in the figure along with the position of 
 \cxou\ and the elliptical boundary of the \hi\ hole.
The best-fit age, mass, and extinction derived for these clusters
 are listed in Table~\ref{t:ic2574sfrsed}.
Also listed are the current mechanical luminosities
 from stellar winds and supernovae, $L_{\rm W}$, corresponding to the
 best-fitting Starburst99 model,
 X-ray background-subtracted net counts in circular apertures of size equal to the 99\% FUV encircled energy
 region in the 0.5$-$2.0~keV energy range,
  and the corresponding X-ray luminosity, $L_{\rm X}$,
 in the 0.5$-$2.0~keV energy range assuming a 0.5~keV thermal plasma X-ray spectral shape.

All clusters are young and
 the masses of the clusters are
 on the order of 10$^{4}$~\msun\ with the exception of C2.
C2 is the central stellar cluster located within the \hi\ hole
 surrounded by the supergiant shell.
The age of this cluster has been estimated as 11~Myr from the FUV
 observation \citep{stewart00} and 5~Myr from the optical
 data \citep{pasquali08}. We obtain a slightly older age.
\citet{stewart00} also estimate the mass of the cluster as
 1.4$\times$10$^{5}$~\msun\
 consistent with the present result of 1.5$\times$10$^{5}$~\msun.
Both authors also found that the \hii\ regions in the vicinity of
 the cluster are younger, having ages of 1$-$3~Myr, which is less
 than the present results.
(We note that we obtain ages of 2$-$3~Myr if we assume a canonical
 dust temperature of 70~K.)
This age gradient supports the premise of propagating star formation
 such that the mechanical energy from the central cluster
 compresses the surrounding ISM and triggers subsequent 
 star formation at the rim.
The mass of C2 is large enough
 to drive star formation in the
 surroundings:
The derived mechanical luminosity for C2 is 2.9 $\times$
 10$^{39}$~erg~s$^{-1}$.  
The cumulative energy released 
 is 2$\times$ 10$^{54}$~erg over the lifetime of the cluster, 
 which is much more than needed to
 create the hole
 \citep[2 $\times$ 10$^{53}$~erg;][]{walter98}.

\section{Summary and Discussion} \label{s:discuss}

High-angular resolution \cxo\ spectrophotometric imaging of the \hi\ hole 
 and surrounding \sgs\ in \ic\ reveal a bright point source,
 \cxou, near the center of the \hi\ hole and residual
 low surface brightness soft X-ray emission throughout the hole and in
 the surrounding shell.
Analysis of the UV-to-IR spectral energy distributions of several
 representative star clusters in and around the \hih\ results in
 age and mass estimates consistent with previous studies.
In particular, there is a massive, 1.5$\times$10$^5$~\msun,
 intermediate-age, 5$-$17~Myr, cluster near the center of the hole
 that has released enough mechanical energy from massive stars and
 supernovae over its lifetime to have created the hole. 
Our analysis is  consistent  with previous analyses of the \sgs\ in \ic\
 \citep{walter98,walter99,stewart00,pasquali08,weisz09} and
supports the classical picture of bubble formation and evolution such that
 younger star clusters on the periphery of the hole likely formed as 
 a consequence of this stellar feedback sweeping ambient material
 into dense clouds
 that collapse under self-gravity and subsequently form new stars
 \citep[e.g.,][]{larson74,weaver77,mccray87,maclow88}.
  
We have shown that the ratio of X-ray to mechanical luminosity
 in these young star clusters in the region is $\lesssim$ 0.2\%.
Observed ratios of 0.1\% to 1\% are typical of
 many star-forming regions \citep{yukitaD}
 but are poorly constrained theoretically because of uncertainties
  in ambient conditions, starburst histories, and three-dimensional geometries.
For example, we note that the central cluster alone could be responsible for all
the X-ray diffuse emission within the \hih;
the mechanical luminosity from the central cluster is
 2.9$\times$10$^{39}$~\ergl\ and the diffuse X-ray luminosity from the hole is
 2.7$\times$10$^{37}$~\ergl.
\citet{weisz09} estimate that the cumulative energy released from the
 whole region in the past $\sim$25~Myr is $\sim$10$^{55}$~erg.
Even at a constant (current) rate,
 less than 0.1\% of the total energy from stars
 has been lost to the X-ray radiation.
Thus, the central cluster in the \hih\ alone can account for the
 diffuse X-ray emission and the creation of the \hih.
Perhaps we are witnessing powerful positive feedback from this central
 cluster in the form of subsequent star formation in the
surrounding \sgs.

Most of the X-ray 
emission in the \hih\ in \ic\ comes from the single bright source \cxou.
The presence of such a luminous X-ray point source in even a 
 massive \sgs\ is uncommon.
Recent supernovae can be this luminous but they, too, are rare
 and typically much softer than \cxou.
The high luminosity, hard spectral shape, and 
 large numbers of recently-formed stars in the vicinity suggest that
 \cxou\ is an HMXB.
We can estimate the age of the binary system at between 
 5~Myr, the youngest age of the nearby massive star cluster, and
 25~Myr, the age of the peak in the star-formation rate
 at the location of \cxou\ as determined by \citet{weisz09}.
The compact object must then have had a minimum initial mass between
 about 10 and 40~\msun\ to have already evolved to a supernova endpoint.
The companion is likely also of roughly this mass in order to
 provide the required high mass accretion rate.
This rate can be estimated assuming 
 $L_{\rm X} \sim L_{\rm bol} \sim \eta \dot{m} c^2$ 
 with an efficiency, $\eta \sim 0.1$, giving 
 $\dot{m} \sim 10^{-7}$~\msfr.
According to \citet{prh03}, this requires a companion star mass
 of at least 8$-$10~\msun\ and higher if the system
 is older, if the bolometric luminosity exceeds the X-ray luminosity,
 or if mass is lost from the system.
\citet{prh03} also estimate the rate of formation of 
 systems of this type scaled to a fiducial supernova rate.
Their rate can be converted to a total number of systems
 in the central star cluster as there are about one $M>8$~\msun\
 star (i.e., stars that eventually undergo supernovae) per 50~\msun\
 of star formation for a Salpeter IMF. 
Thus, the \citet{prh03} rate of 
 $<$10$^{-4}$~yr$^{-1}$ per 8~\msun\ star translates to 
 0.3 luminous HMXBs for the central star cluster of mass 1.5$\times$10$^5$~\msun.
This is a high estimate because not all the potential compact objects have
 yet been created by supernova events. 
Thus, objects like \cxou\ are indeed rare even for massive star clusters.

We gratefully acknowledge the  referee, Leisa Townsley, for her expert critique and
 especially for alerting us to the possibility of pile-up of the point source.
 Support for this research was provided in part by NASA through an Astrophysics Data Analysis
 Program grant NNX08AJ49G.

\end{document}